\documentclass[aps,showpacs]{revtex4}
\usepackage{graphicx}
\usepackage{dcolumn}
\usepackage{bm}
\newcommand{\bsigma}{\mbox{\boldmath $\sigma$}}
\newcommand{\btau}{\mbox{\boldmath $\tau$}}

\newcommand{\br}{{\bf r}}

\newcommand{\car}{$^{12}$C}
\newcommand{\oxy}{$^{16}$O}
\newcommand{\caI}{$^{40}$Ca}
\newcommand{\bqu}{{\bf q}}
\newcommand{\bpi}{{\bf p}}
\newcommand{\e}[1]{ {\rm e}^{#1} }
\newcommand{\bkey}{{\bf k}}
\newcommand{\polga}{\vec{\gamma}}
\begin{document}
\title {Proton emission induced by polarized photons}
\author {M. Anguiano$ ^{\,1}$, G. Co'$ ^{\,\,2,3}$ and
A.M. Lallena$^{\,1}$} 
\affiliation {$^{1)}$ Departamento de F\'{\i}sica
At\'omica, Molecular y Nuclear, Universidad de Granada, 
E-18071 Granada, Spain \\
$^{2)}$ Dipartimento di Fisica,  Universit\`a di Lecce, 
I-73100 Lecce, Italy \\
$^{3)}$  Istituto Nazionale di Fisica Nucleare  sez. di Lecce,
 I-73100 Lecce, Italy
} 

\date{\today}

\begin{abstract}
  The proton emission induced by polarized photons is studied in the
  energy range above the giant resonance region and
  below the pion emission threshold.  Results for the \car,
  \oxy \/ and \caI \/ nuclei are presented. The sensitivity of
  various observables to final state interaction, meson exchange
  currents and short range correlations is analyzed. We found relevant
  effects due to the virtual excitation of the $\Delta$ resonance.
\end{abstract} 
\maketitle
%
\section{INTRODUCTION}
\label{sec:intro}

In a previous work \cite{ang02} we have investigated the sensitivity
of the $(\gamma,p)$ reactions to short range correlations (SRC),
meson-exchange currents (MEC) and final state interactions (FSI).  The
SRC are related to the short-distance repulsion of the nucleon-nucleon
potential \cite{arn92,wir95}. The search for SRC effects in nuclear
systems is done by studying deviations from mean field results
\cite{ben93,pan97}.  The exchange of mesons between interacting nucleons
produces electromagnetic currents called MEC.  Effects related to the
MEC have been clearly identified in few-body systems
\cite{car98,mar05}.  In heavier nuclei, however, a clean
identification of MEC effects is hidden, to a large extent, by the
uncertainties of the nuclear wave function \cite{bof84}.  The FSI
account for the re-interaction of the emitted nucleon with the
remaining nucleus.  A successful approach to describe nucleon
emission data induced by both electrons and photons
\cite{bof85,bof93}, treats the FSI with an optical potential whose
parameters are fixed to reproduce elastic nucleon-nucleus scattering
data.

Almost all the experimental work relative to the photo-emission of a
single proton in medium-heavy nuclei, has been done with unpolarized
photons \cite{bof96}. We are aware of only two experiments of
$(\polga,p)$ type.  In the first one, Wienhard {\it et al.}
\cite{wie81} studied the $^{16}$O($\polga,p$)$^{15}$N reactions in the
giant resonance region by using photons with energies between 15 and
25 MeV.  In the other experiment, Yokokawa {\it et al.} \cite{yok88}
used as a target the $^{12}$C nucleus and the photon energies varied
between 40 and 70 MeV.

The recent development of tagged photon facilities opens new
perspectives for this type of experiments.  For example the Mainz
Microtron (MAMI) \cite{mam06w} and the new tagged photon beam line at
MAX-laboratory, in Lund \cite{max06w}, can make experiments with
polarized photons and produce data with a high energy resolution. This
allows a clear separation of the different states of the residual
nucleus \cite{mil95}.  At MAMI, polarized photon beams have been
already used to study $^{12}$C($\polga,pd$), $^{12}$C($\polga,pp$) and
$^{12}$C($\polga,pn$) reactions \cite{fra99}.  At MAX-laboratory two
experiments with polarized photons have been proposed.  In the first one, 
linearly polarized bremsstrahlung photons will be used to
make Compton scattering on $^{4}$He and \car \/ nuclei. In the
second experiment cross sections and asymmetries of the two-particle
photodisintegration of $^{6}$Li and $^{7}$Li nuclei will be
measured.  Another experimental facility which produces polarized
gamma-ray beams is HIGS, at the Duke University in Durham
\cite{hig06w}, where the first measurements of the
$^{2}$H($\vec{\gamma}$,n)p analyzing power near threshold have been
done \cite{sch00}.

The experimental situation is rapidly evolving and the possibility of
using polarized photons to study the structure of medium-heavy nuclei
is a solid perspective. In this work we study this possibility from
the theoretical point of view, by using a nuclear model recently
developed to investigate electromagnetic excitations of the nucleus in
inclusive \cite{co98,ama98,mok00,co01}, single \cite{mok01,ang02} and
double coincidence \cite{ang03,ang04} experiments.  The starting point
of our approach is the continuum shell model implemented with the
optical potential to take into account the FSI. By using this model to
describe nuclear excited states we treat the MEC by considering
one-pion exchange diagrams \cite{ama93,ama94,ama93t}.  We improve this
picture by implementing the SRC acting on one-body electromagnetic
currents.  The SRC are considered by calculating all the diagrams
containing a single correlation line. The validity of this approach
has been verified in Ref. \cite{ama98} by comparing nuclear matter
charge responses obtained with our model and with a full Correlated
Basis Function calculation.

In our previous investigation \cite{ang02} we found that photoemission
cross sections have a greater sensitivity to SRC than electron
scattering cross sections.  The SRC effects are relevant at photon
energies above 100 MeV and for proton emission angles above 80 degrees.
Unfortunately, in these kinematics, also the MEC play an important
role, and their effects hide those of the SRC. For these reasons we
find interesting to investigate the possibility of disentangling MEC
and SRC with polarization observables.

The paper is organized as follows. In Sect. \ref{sec:model} we briefly
define the observables we want to study, and present the basic ideas of
our model. In Sect. \ref{sec:results} we show our results and
discuss separately the effects of the FSI, of the SRC and of the MEC.
Finally, in Sect. \ref{sec:conclusion} we summarize our findings and
draw our conclusions.

\begin{figure}[b]
\begin{center}
\parbox[c]{16cm}
{\includegraphics[scale=0.7]{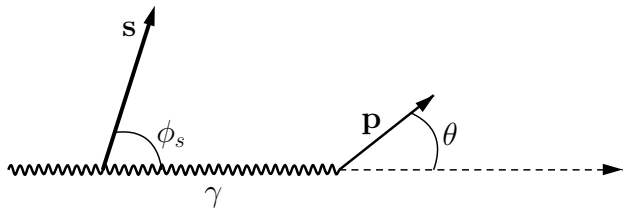}}
\end{center}
\vspace*{-.6cm}
\caption{\small Geometry of the process. 
  We call $\phi_s$ the angle between the directions of the spin and of
  the momentum of the photon.  The angle $\theta$ defines
  the proton emission direction with respect to the photon momentum.
}
\label{fig:axis}
\end{figure}

\section{THE NUCLEAR MODEL}
\label{sec:model}
We describe a process where a proton of momentum $\bpi$ is emitted
from a nucleus after the absorption of a linearly polarized photon of
momentum $\bqu$ and energy $\omega$.  The geometry of the process is
presented in Fig. \ref{fig:axis}. We indicate with $\phi_s$ the angle
between the spin and the momentum of the photon and with $\theta$ the
angle between the $\bqu$ and $\bpi$ vectors. The cross section for
this process can be written as \cite{bof96,ang02}:
\begin{equation}
\frac {d \sigma (\omega,\phi_s,\bpi)} {d \theta} =
\frac{2 \pi^2 e^2 } {\omega}
\frac{|\bpi| m_p} {(2 \pi)^3}
\left[ W_{\rm T}(\omega,\omega,\bpi) + W_{\rm TT} (\omega,\omega,\bpi)
\cos \phi_s  \right] \, ,
\label{eq:xsect}
\end{equation}
where $m_p$ is the proton mass and the nuclear responses 
$W_{\rm T}$ and $W_{\rm TT}$ are defined as:
\begin{eqnarray}
\mbox{W}_{\rm T} (q, \omega, \bpi) & = & \sum_{\eta = \pm 1} |\ \langle
{\Psi}_f (\bpi)| J_{\eta}(q)| {\Psi}_i \rangle |^2 
\delta (E_f-E_i-\omega) \, ,
\label{eq:wt}
\\ 
\mbox{W}_{\rm TT} (q,\omega, \bpi) & = & 2 Re \left ( 
\langle \Psi_i |
J_{+1}^{\dagger} (q)| \Psi_f (\bpi) \rangle \langle \Psi_f (\bpi) 
|J_{-1} (q)
|\Psi_i \rangle \right ) \delta (E_f-E_i-\omega) \, .
\label{eq:wtt}
\end{eqnarray} 
In the above equations $|\Psi_i \rangle$ and $|\Psi_f (\bpi) \rangle$
indicate the initial and final states of the nuclear system, with
energies $E_i$ and $E_f$ respectively. Since we deal with real photons
we have that, in natural units, 
$\omega = E_f-E_i = |\bqu| \equiv q$, and only
the transverse components of the nuclear current  $J_{\pm 1}$ are
active:
\begin{eqnarray}
J_{\pm 1} = \mp \frac{1}{\sqrt{2}} (J_x \pm i J_y) \, .
\end{eqnarray}
In electron scattering there is an additional contribution to the
electromagnetic current, a longitudinal term, related to the nuclear
charge distribution.

In our calculations we have considered only the situations where the
spin of the photon is parallel or orthogonal to $\bqu$. We call
respectively $\sigma_{\parallel}$ and $\sigma_{\perp}$ the cross
sections of these two cases. We have studied the photon asymmetry
defined as \cite{bof96}:
\begin{eqnarray}
\Sigma =
\frac{\sigma_{\parallel}-\sigma_{\perp}}
{\sigma_{\parallel}+\sigma_{\perp}}=
\frac{W_{\rm TT}}{W_{\rm T}} \, .
\label{eq:sigma}
\end{eqnarray}

A linearly polarized photon gives information on the structure
function W$_{\rm TT}$, which does not contribute to the unpolarized
cross section.  The asymmetry $\Sigma$ can be thought as the
correction factor needed to obtain the polarized cross section
($\sigma_{\rm pol}$) from the unpolarized one ($\sigma_{\rm unpol}$),
i.e,
\begin{eqnarray}
\sigma_{\rm pol}=\sigma_{\rm unpol} \, \left (1+\Sigma \, 
\cos \, 2\phi_s \right ) \, .
\end{eqnarray}
Since $W_{\rm T}$ is always positive, the sign of the
asymmetry is the sign of the structure function $W_{\rm TT}$.

We have restricted our calculations to doubly closed shell nuclei;
then, the nuclear ground state has zero angular momentum and positive
parity, i.e, $|\Psi_i\rangle=|J_i M_i; \Pi_i \rangle = |00; +1
\rangle$.  We describe the ground state as a Slater determinant of
single-particle wave functions produced by a mean-field potential of
Woods-Saxon type with the parameters given in Ref. \cite{bot05} for
\car \/ and Ref. \cite{ari96} for \oxy \/ and \caI.  We write the
nuclear final state as \cite{co85}:
\begin{eqnarray}
|\Psi_f \rangle & = & \frac{4 \pi} {|p|} \sum_{l_p \mu_p}
\sum_{j_pm_p} \sum_{JM, \Pi} i^{l_p} Y_{l_p \mu_p} (\hat{p}) \langle
l_p \mu_p \frac{1}{2} \sigma | j_p m_p \rangle \nonumber \\ & \times &
\langle j_p m_p j_h m_h | J M \rangle | \Psi; J M ; \Pi; (l_p j_p m_p
\epsilon_p, l_h j_h m_h \epsilon_h ) \rangle \, .
\end{eqnarray}
In the above equation, $|J M;\Pi;(l_p j_p m_p \epsilon_p, l_h j_h
m_h \epsilon_h ) \rangle $ describes the excited state of the $A$
nucleons system with total angular momentum $J$, $z-$axis projection
$M$, and parity $\Pi$. This state is composed by a particle in a
continuum wave, with orbital and total angular momenta $l_p$ and $j_p$
respectively, projection $m_p$, energy $\epsilon_p$ and momentum ${\bf
  p}$, and a residual nucleus with hole quantum numbers $l_h$, $j_h$,
$m_h$ and $\epsilon_h$.  We have indicated with $Y_{l\mu}$ the
spherical harmonics and with $\langle l_am_al_bm_b|JM \rangle$ the
Clebsch-Gordan coefficients \cite{edm57}.

In our calculations the $|J M;\Pi;(l_p j_p m_p \epsilon_p, l_h
j_h m_h \epsilon_h ) \rangle $ state is a Slater determinant
constructed on the ground state Slater determinant by substituting the
hole wave function $ |\phi_h \rangle $ with the continuum wave
function $ | \phi_p \rangle $.  This continuum single particle wave
function is calculated by using a complex optical potential which is
supposed to take into account the FSI. With this procedure, particle
and hole wave functions are not  any more orthogonal. The effects of
this inconsistency have been found to be negligible in the kinematics
under investigation \cite{bof82}.

The continuum shell model above presented has been modified to take
into account the SRC in both initial and final state. We have
considered only the case of scalar correlation functions acting on
one-body (OB) currents. Following the basic steps of the Correlated 
Basis Function theory, we made a cluster expansion of the transition
matrix elements of Eqs. (\ref{eq:wt}) and (\ref{eq:wtt}) to eliminate
the unlinked diagrams \cite{fan87}. At this point we restrict our
calculations by considering all, and only, the terms linear in the
correlation function. This implies the evaluation of four two-body
diagrams and six three-body diagrams, for each OB operator considered.
This procedure is necessary to guarantee the correct normalization of
the many-body wave functions \cite{co95}. A detailed description of
the SRC model can be found in \cite{co01,mok01,ang02}.

\begin{figure}[b]
\begin{center}
\parbox[c]{16cm}{\includegraphics[scale=0.8]{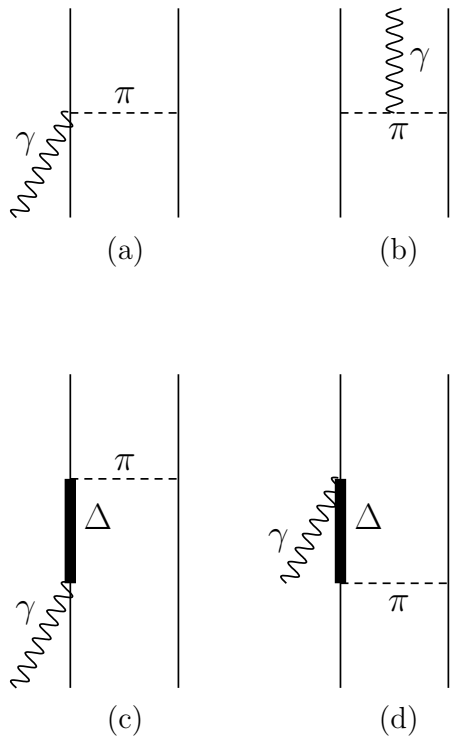}}
\end{center}
\vspace*{-.6cm}
\caption{\small Feynmann diagrams of the MEC terms considered
  in our calculations. The (a) and (b) diagrams represent the 
  seagull and pionic currents, respectively, while the other two  
  diagrams the $\Delta$ currents.
}
\label{fig:diamec}
\end{figure}

In our calculations we have 
considered only the MEC terms presented in Fig.
\ref{fig:diamec} generated by the exchange of a single pion. Following
the  model
developed in Refs. \cite{ama93,ama94,ama93t} 
we have calculated the seagull
diagram (A in Fig. \ref{fig:diamec}), the pionic diagram 
(B in Fig. \ref{fig:diamec}) and the other two $\Delta$ currents 
diagrams.
The expressions of the seagull and pionic terms are given in 
Ref. \cite{ang02}. For the $\Delta$ currents terms 
we use the following expression, more general than that used in
\cite{ang02}:
\begin{eqnarray} 
\nonumber {\bf j}^{\Delta}({\bf q},\omega) &= & -i
\frac{2}{9}\frac{f_{\pi N \Delta} f_{\pi N N} f_{\gamma N \Delta}}
{m_{\pi}^3} \sum_{k,l=1 \atop k\ne l}^A \, \e{i\bqu\cdot\br_k}\bqu
\cdot \\ & &
\displaystyle \left\{ 
[ \btau (k) \times \btau (l)]_3 \,
 \bsigma (k) \times \nabla_k \, 
 \bsigma (l) \cdot \nabla_k \, h(\varepsilon_{kl},{\bf r}_k-{\bf r}_l) 
\right. \\ & &  \nonumber \left.  - \, 4 \tau^3 (l) \, \nabla_k \, 
 \bsigma (l) \cdot \nabla_k \, h(\varepsilon_{kl},{\bf r}_k-{\bf r}_l)
\right\}  
+ (  k \leftrightarrow l ) \, . 
\label{eq:isobar}
\end{eqnarray}
where $f^2_{\pi N N} / 4 \pi = f^2_{\pi} = 0.079$ is the effective
pion-nucleon coupling constant, $m_{\pi}$ is the pion mass and
$h(\br)$ is the Fourier transform of the dynamical pion propagator
\cite{ama94}
\begin{equation}
h(\epsilon,\br-\br_l)=\int \frac{{\rm d}^3k}{(2\pi)^3}\,
\frac{F_{\pi\rm N}(k,\varepsilon)\, 
\e{i\bkey \cdot (\br-\br_l)}} 
{k^2+m_{\pi}^2-\varepsilon^2} \, .
\label{eq:prop}
\end{equation}

We have indicated with $F_{\pi\rm N}$ the pion-nucleon form factor and
with $\varepsilon$ the energy of the exchanged pion.  As it is clear
from Fig. \ref{fig:diamec} MEC are two-body operators, then they could
lead also to excited state with two particle in the continuum.  We
neglect their contribution which we found to be extremely small in
inclusive processes \cite{co01}, and we expect to be negligible also
in the case under investigation. In our calculations the four single
particle states involved in the MEC calculations are, the continuum
state of the emitted particle, the hole state characterizing the
residual nucleus, and other two states below the Fermi level.  In
this situation the $\varepsilon$ energies of Eq. (\ref{eq:isobar}) are
uniquely defined.  As stated above, in our model the SRC act only on
the OB currents, therefore MEC and correlations interplay only trough
the interference between the transition amplitudes.

\section{RESULTS}
\label{sec:results} 

We present in this section the results of the $(\polga,p)$ reaction on
\car, \oxy \/ and \caI \/ target nuclei. We are interested in the excitation
energy region above the giant resonance and below the pion production
threshold. In this region, the collective phenomena characterizing the
giant resonances are not any more relevant \cite{ang02}, and the
internal structure of the nucleon is easily parametrized. For these
reasons it is plausible to attribute the corrections to a mean-field
description of the process to the effects we want to study: FSI, SRC
and MEC. The relevance of these effects on the asymmetry will be
investigated by considering them separately. In a second step we shall
present the results obtained by putting all ingredients together.

\subsection{Final state interactions}
\label{sec:FSI}

We have calculated the asymmetries by using different optical
potentials to describe the emitted proton wave function. In our
calculations we used the potentials of Schwandt {\it et al.} (Sc)
\cite{sch82}, of Comfort and Karp (CK) \cite{com80} and of Abdul-Jalil
and Jackson (AJ) \cite{abd80}.  In addition, we have done calculations
by using, also for the particle states, the same real Woods-Saxon
potential considered for the hole states.  We label the results of
these last calculations as WS.

\begin{figure}[ht]
\begin{center}
{\includegraphics[width=10.3cm]{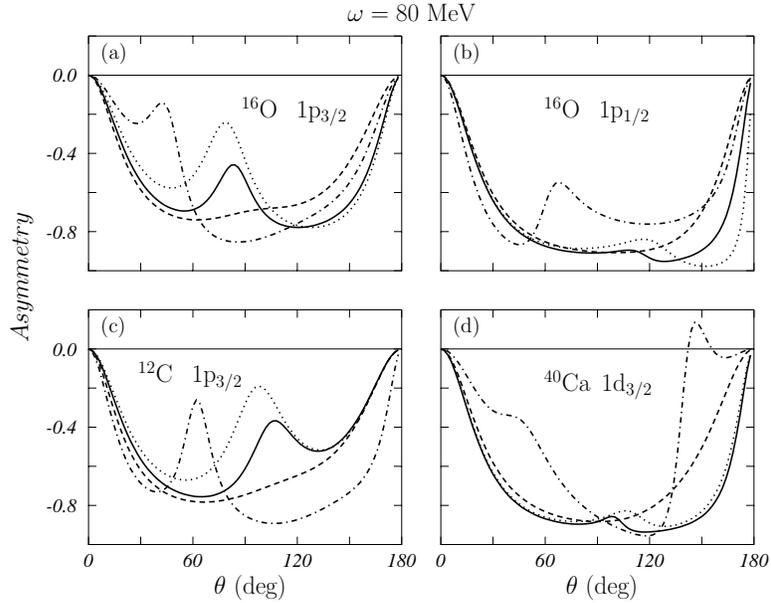}}
\end{center}
\vspace*{-.6cm}
\caption{\small Asymmetries of the $(\polga,p)$ reaction, 
  eq. (\protect\ref{eq:sigma}), calculated by using OB currents only.
  The target nuclei and the hole states of the remaining nuclei are
  indicated in each panel. The photon energy has been fixed at
  $\omega=80$ MeV. The lines show the results obtained with various
  optical potentials. Specifically, the solid lines have been obtained
  with the Sc potential \protect\cite{sch82}, the dotted lines with
  the CK potential \protect\cite{com80}, the dashed lines with the AJ
  potential \protect\cite{abd80}, and the dashed-dotted lines with the 
  real WS potential \protect\cite{ari96,bot05}.
  }
\label{fig:OBOP}
\end{figure}

In Fig. \ref{fig:OBOP} we show the asymmetry as a function of the
proton emission angle. We fixed the photon energy at 80 MeV and we
consider the \car, \oxy \/ and \caI \/ nuclei. In each panel of the 
figure the hole state of the remaining nucleus is indicated. These
calculations have been done by considering OB currents only, and
without SRC. The difference between the various lines is due only to
the use of different optical potentials.

The first remark is that all the asymmetries have the same order of
magnitude and show similar behaviors.  The differences between the
various results are in the detailed structure of the angular
distributions. We observe that the results obtained with the Sc and
the CK optical potentials (solid and dotted lines respectively) have
very similar behaviors, and show peaks roughly located at the same
angle.  On the contrary, the results obtained with the AJ potential
(dashed curves) do not show any peak at all.  Finally, the angular
distributions of the WS results (dashed-dotted curves) are rather
different from all the other ones.

\begin{figure}[ht]
\begin{center}
{\includegraphics[width=10.3cm]{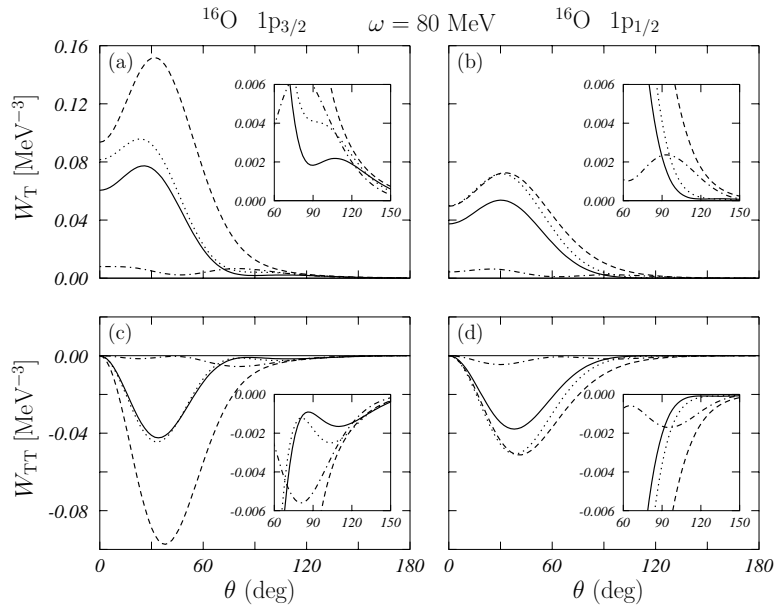}}
\end{center}
\vspace*{-.6cm}
\caption{\small The W$_{\rm T}$ (upper panels) and W$_{\rm TT}$ (lower
panels) response functions contributing to the asymmetries for the
emission from the 1p$_{3/2}$ (left panels) and 1p$_{1/2}$ (right
panels) hole states. The meaning of the lines is the same as in
Fig. \ref{fig:OBOP}. In the insert the behaviour of the responses for
$\theta \ge 60^o$ has been emphasized.}
\label{fig:16O-OBOP-wtwtt}
\end{figure}

To have a better understanding of these results, we show in Fig.
\ref{fig:16O-OBOP-wtwtt} the W$_{\rm T}$ and W$_{\rm TT}$ response
functions for the cases of the 1p$_{3/2}$ and 1p$_{1/2}$ hole states
in \oxy, corresponding to the panels (a) and (b) of Fig.
\ref{fig:OBOP}. Contrary to what happens for the asymmetries, the
responses have rather different sizes.  For example, the WS responses
are remarkably smaller than the other ones. This effect has been
already discussed in \cite{ang02} where we have shown that the
($\gamma,p$) cross sections become smaller the stronger is the real
part of the particle mean-field potential.

The shape of angular distributions of the asymmetries is
characteristic of the hole state angular momentum. In Fig.
\ref{fig:OBOP} the results obtained for the 1p$_{3/2}$ state of \oxy
\/ (panel (a)) and \car \/ (panel (c)) nuclei, have similar
structures.  They are very different from those of the 1p$_{1/2}$
state for \oxy \/ (panel (b)). We found similar results for the \caI
\/ hole states.  Also in this case the study of the responses helps in
understanding the behavior of the asymmetries. We observe in Fig.
\ref{fig:16O-OBOP-wtwtt} that the angular distributions of the
responses obtained with complex optical potentials, are rather
similar. The sizes of these responses are about the same for Sc and
CK. On the contrary, the AJ potential produces responses that for the
1p$_{3/2}$ case are almost two times larger than those of the
1p$_{1/2}$ case. We found similar results for the other photon
energies we have investigated.

For the 1p$_{3/2}$ hole, the CK and Sc potentials provide W$_{\rm T}$
and W$_{\rm TT}$ responses which show a local minimum and maximum,
respectively, around 90 degrees (see inserts in panels (a) and
(c)). This structure is responsible for the peak observed at this
emission angle in the asymmetry (see panel (a) in
Fig.\ref{fig:OBOP}). The situation is different for the 1p$_{1/2}$
state and the asymmetry in this case (see panel (b) in Fig.\ref{fig:OBOP})
is flat for emission angles between 60 and 150 degrees.

We can conclude that the asymmetry is less sensitive to the details of
the optical potential than the cross section. The optical potential
slightly modifies the shapes of the angular distributions of the
asymmetries, but their sizes, and their general behaviors are almost
independent from the choice of the optical potential.

The study of the ($\gamma,p$) cross sections \cite{ang02} indicates
that, in the excitation energy region we want to investigate, the Sc
and CK potential provide a good description of the experimental data.
These two potentials produce very similar results. For these reasons,
henceforth, if not explicitly mentioned, we shall present results
obtained with the Sc optical potential.

\begin{figure}[b]
\begin{center}
{\includegraphics[width=10.3cm]{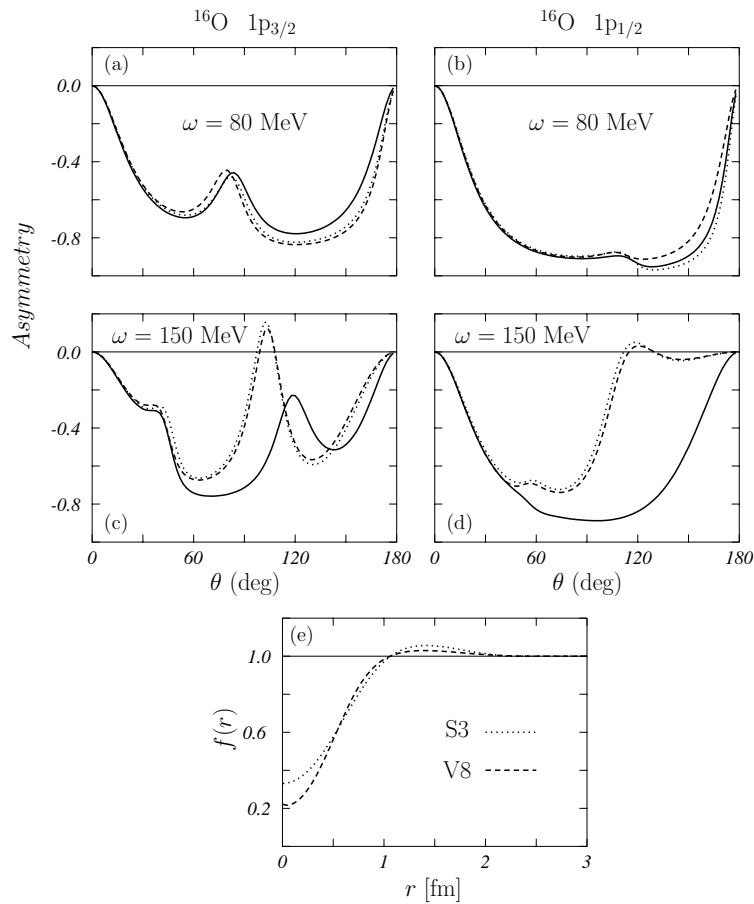}}
\end{center}
\vspace*{-.6cm}
\caption{\small Angular distribution of the asymmetries for excitation
  energies of 80 and 150 MeV. These results have been obtained by
  using OB currents only. The full lines show the results obtained
  without the SRC  and they are the same as in Fig. \ref{fig:OBOP}.
  Dashed and dotted lines have been obtained with the correlations
  shown in panel (e), taken from Ref. \protect\cite{ari96} (dashed
  lines) and from Ref. \cite{bis06} (dotted lines).  }
\label{fig:SRC}
\end{figure}

\subsection{Short range correlations}
\label{sec:SRC}

In this section we study the sensitivity of the asymmetries to the
SRC. In our previous study of the ($\gamma,p$) processes \cite{ang02},
we found that the SRC effects become important at energies above 100
MeV and for large values of the proton emission angle. For this reason
we show in Fig. \ref{fig:SRC} asymmetries calculated with photon energies
of 80 (panels (a) and (b)) and also of 150 MeV (panels (c) and (d)).

The results shown in Fig. \ref{fig:SRC} have been obtained by using OB
currents only. The full lines have been obtained without SRC.  The
other lines show the results obtained by using the scalar term of
correlation functions taken from Fermi Hypernetted Chain calculations
for finite nuclear systems. The dotted lines of Fig.  \ref{fig:SRC}
have been obtained with the correlation of Ref.  \cite{ari96} fixed to
minimize the energy obtained with the semirealistic S3
nucleon-nucleon interaction of Ref. \cite{afn68}.  The dashed lines
have been obtained by using the scalar term of the state dependent
correlation function of Ref. \cite{bis06}. In these last calculations
the realistic Argonne $V8'$ interaction \cite{pud95} has been used.
The two correlations used are shown in the panel (e) of the figure.

In agreement with Ref. \cite{ang02}, the effects of the SRC on the
asymmetries become remarkable at high energies and for large values of
the emission angle.  The sensitivity of the probe is however not
sufficient to disentangle the two correlation functions.

\subsection{Meson exchange currents}
\label{MEC}

We investigate the effects of the MEC by separating the contribution
of the seagull and pionic terms, corresponding to the (a) and (b)
diagrams of Fig. \ref{fig:diamec}, from those of the $\Delta$
currents. The reason of this separation is that while the coupling
constants related to seagull and pionic currents are well determined
by pion-nucleon scattering data, the values of the $f_{\pi N \Delta }$
and $f_{\gamma N \Delta}$ constants, necessary to the evaluation of
the $\Delta$ currents, see Eq. \ref{eq:isobar}, are still slightly
uncertain. Following the procedure adopted in \cite{ang03} we compare
the results obtained with the three different sets of values taken
from Refs. \cite{ama94} (AMA), \cite{giu97} (GIU) and \cite{ryc97}
(RYC) and shown in Table \ref{tab:delta}.
\begin{table}[t]
\begin{center}
\begin{tabular}{rrrr}
\hline\hline
                         &   AMA &   GIU & RYC  \\
\hline
 $f_{\gamma N \Delta}$   & 0.299 & 0.373 & 0.120  \\
 $f_{\pi N \Delta}$      & 1.69  & 2.15  & 2.15  \\
\hline\hline
\end{tabular}
\end{center} 
\caption{\small Values of the parameters used in
Eq. (\protect\ref{eq:isobar}) for the three parameterizations
considered in this work.  The AMA, GIU and RYC values are from
Refs. \protect\cite{ama94}, \protect\cite{giu97} and
\protect\cite{ryc97} respectively.  }
\label{tab:delta}
\end{table}

\begin{figure}[ht]
\begin{center}
{\includegraphics[width=10.3cm]{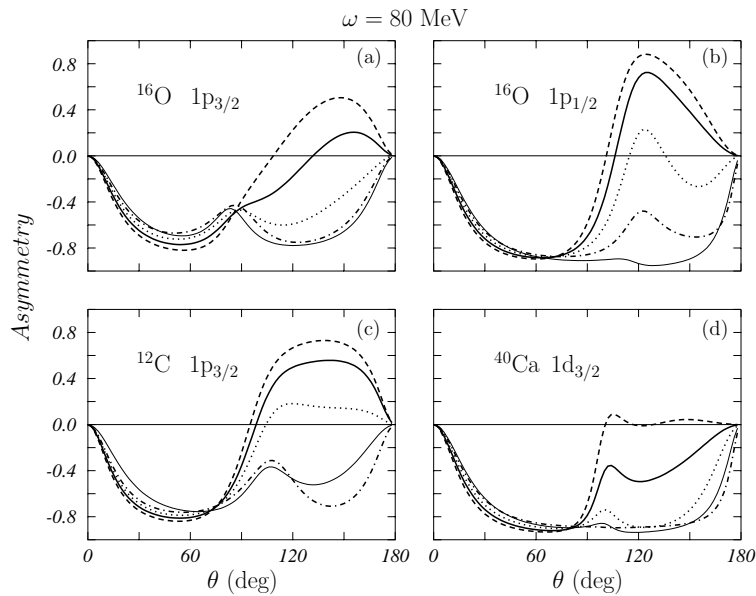}}
\end{center}
\vspace*{-.6cm}
\caption{\small Asymmetry angular distributions.
The results obtained with the OB currrents only, are represented by
the thin solid lines.  The dashed-dotted lines
have been obtained by including the seagull and pionic
currents. 
The other lines show the results obtained by including also
the $\Delta$ currents which have been calculated 
with the coupling constants given in Table \ref{tab:delta}:
AMA (thick full lines), GIU (dashed lines),  RYC (dotted lines).
}
\label{fig:MEC}
\end{figure}

\begin{figure}[ht]
\begin{center}
{\includegraphics[width=10.3cm]{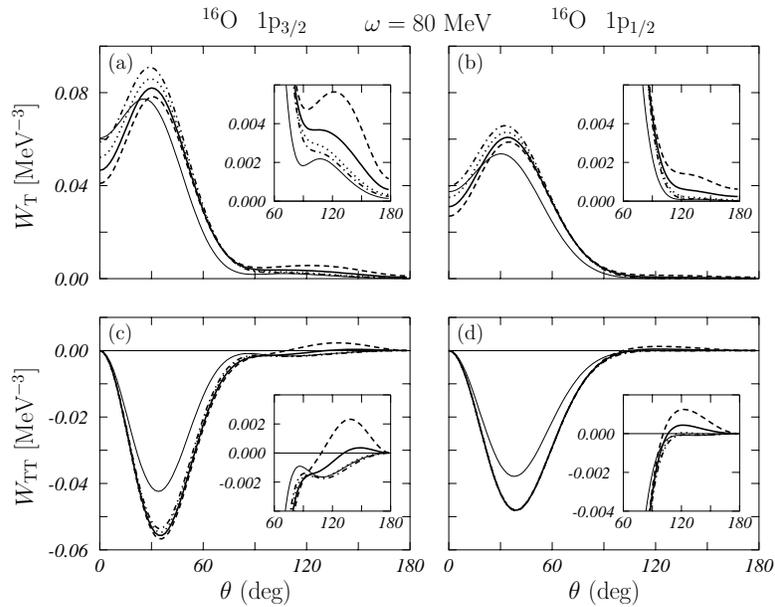}}
\end{center}
\vspace*{-.6cm}
\caption{\small Angular distributions of the 
$W_{\rm T}$ (upper panels) and $W_{\rm TT}$(lower
panels) responses. The meaning of the lines is the same as in Fig. 
\protect\ref{fig:MEC}.}
\label{fig:16O-MEC-wtwtt}
\end{figure}

\begin{figure}[ht]
\begin{center}
{\includegraphics[width=10.3cm]{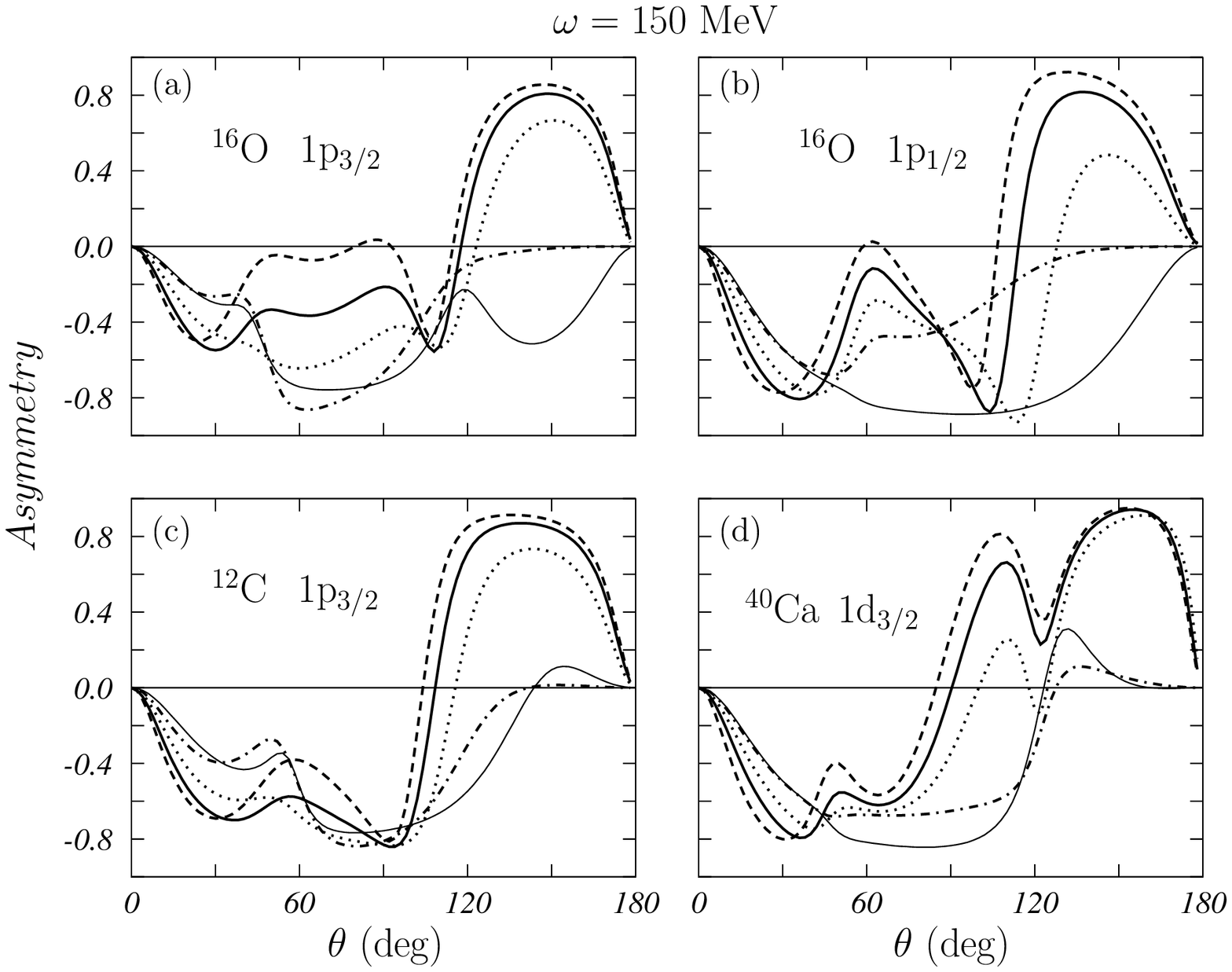}}
\end{center}
\vspace*{-.6cm}
\caption{\small The same as Fig. \protect\ref{fig:MEC} for an
  excitation energy of 150 MeV.}
\label{fig:MEC-higE}
\end{figure}

\begin{figure}[ht]
\begin{center}
{\includegraphics[width=10.3cm]{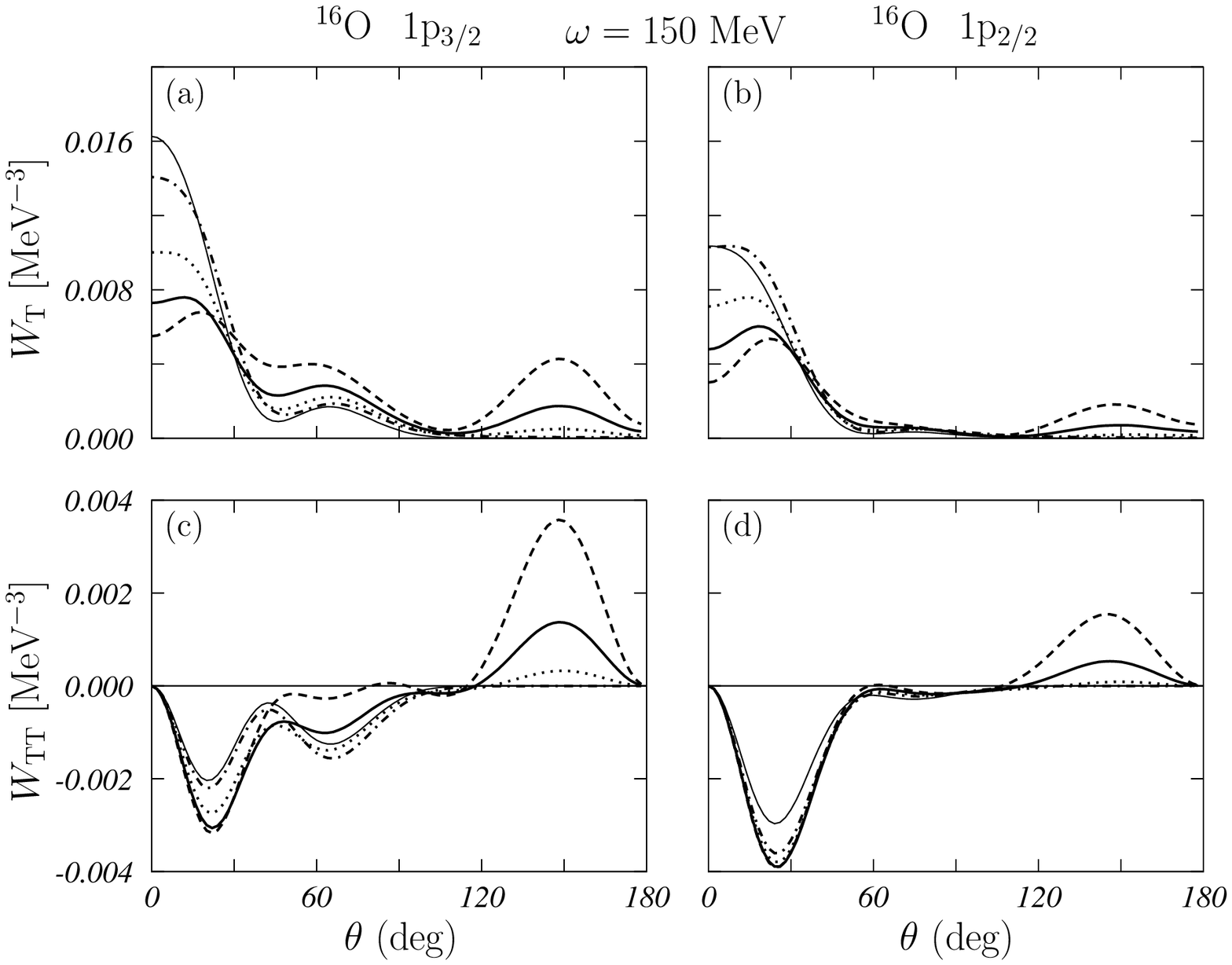}}
\end{center}
\vspace*{-.6cm}
\caption{\small The same as Fig. \protect\ref{fig:16O-MEC-wtwtt} for an
  excitation energy of 150 MeV.}
\label{fig:16O-MEC-wtwtt-higE}
\end{figure}

In Figs.  \ref{fig:MEC}, \ref{fig:16O-MEC-wtwtt}, \ref{fig:MEC-higE}
and \ref{fig:16O-MEC-wtwtt-higE}, we present the angular distributions
of asymmetries and responses calculated at 80 and 150 MeV.  The thin
solid lines show the results obtained with the OB currents only and
the dashed dotted lines those obtained by including the seagull and
pionic terms. The other lines have been obtained by including the
$\Delta$ currents with the three sets of parameters given in Table
\ref{tab:delta}. The full, dashed and dotted lines correspond to the
AMA, GIU, and RYC parameterizations respectively.

The angular distributions of the asymmetry at 80 MeV, see Fig.
\ref{fig:MEC}, show that the seagull and pionic terms of the MEC
produce small effects. On the contrary, the effects of the $\Delta$
currents are remarkable. These effects are very sensitive to the
values of the coupling constants and become more evident at large
emission angles.

The angular distributions of the W$_{\rm T}$ and W$_{\rm TT}$
responses of \oxy \/ at 80 MeV are shown in Fig.
\ref{fig:16O-MEC-wtwtt}.  The MEC currents modify the size of minima
and maxima, but in all the cases the shapes of the angular
distributions are similar. There is an increase of the W$_{\rm T}$
maxima with respect to the pure OB current results.  In any case, the
$\Delta$ currents lower the values obtained by including seagull and
pionic terms only (dashed-dotted curves). The large effects of the MEC
seen in the asymmetries are located in the region of large emission
angle, where the responses are relatively small with respect to the
maximum values found around $\sim 30$ degrees. The behavior of the
responses in this region is emphasized in the inserts of Fig.
\ref{fig:16O-MEC-wtwtt}. We observe that the W$_{\rm TT}$ changes
sign, and this produces the large effects on the asymmetries.

Angular distributions of asymmetries and responses at 150 MeV are
shown in Figs. \ref{fig:MEC-higE} and \ref{fig:16O-MEC-wtwtt-higE}.
The asymmetries show behaviors similar to those seen at 80 MeV. The
two basic facts, small effects of seagull and pionic terms, and large
effects of the $\Delta$ currents, are present also at 150 MeV. We
observe much more complicated angular distribution patterns at low
emission angles, and relatively large effects of  seagull and pionic
terms in \oxy. The $\Delta$ currents dominate the MEC effects, but
their sensitivity to the constant values seems slightly reduced.

The study of the angular distributions of the W$_{\rm T}$ and
W$_{\rm TT}$ terms, Fig. \ref{fig:16O-MEC-wtwtt-higE}, allows a better
understanding of the asymmetries behavior. It is interesting to
observe that the W$_{\rm T}$ responses are peaked in the forward
direction for both 1p$_{1/2}$ and 1p$_{3/2}$ emission.  The inclusion
of the MEC lower the values of these peaks and strongly change the
shape of the distributions. On the contrary, the  W$_{\rm TT}$
responses are not so sensitive to the presence of the MEC. Therefore the 
W$_{\rm T}$ responses are responsible of the large MEC effects we
found in the asymmetries.

\begin{figure}[htb]
\begin{center}
{\includegraphics[width=10.3cm]{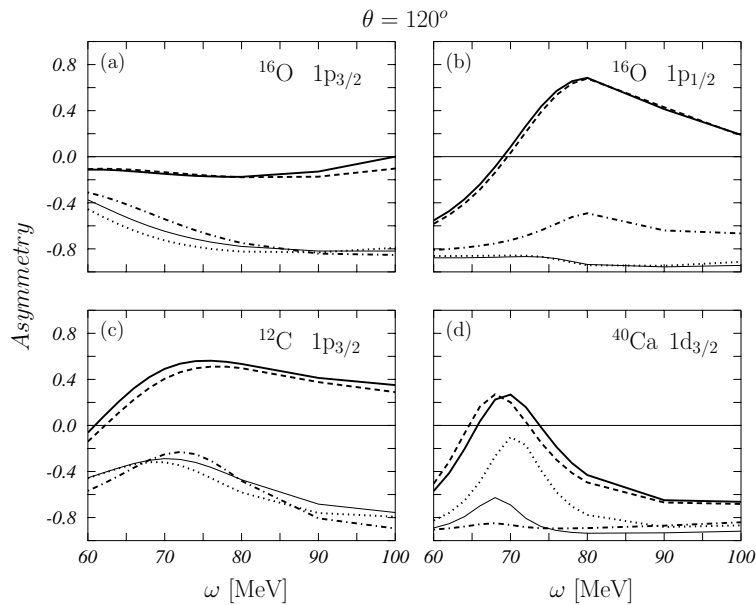}}
\end{center}
\vspace*{-.6cm}
\caption{\small Asymmetries as a function of the excitation energy for
the proton emission angle $\theta=120^o$. The full thin lines have been
obtained with the OB currents only, the dotted lines by adding the
SRC, the dashed lines by including MEC and the thick continous lines
by considering both MEC and SRC. The dashed-dotted lines include OB,
seagull and pionic currents and SRC.}
\label{fig:MEC-theta120}
\end{figure}

The relative effects of MEC and SRC are evident in Fig.
\ref{fig:MEC-theta120} where we show asymmetries calculated at
$\theta=120^o$ as a function of the excitation energy. The thin
solid lines show the results obtained with OB currents only. The
dotted lines show the results when the SRC correlations are
included; the dashed lines have been obtained by including the
MEC; the dashed-dotted lines include all terms except the contribution
of the $\Delta$ current, and the thick solid lines have been obtained
by including both MEC and SRC. The parametrization AMA has been used
for the $\Delta$ current. It is evident that the dominant effect
beyond the mean-field description of the process is that of the
MEC. Coherence effects produced by the interference of MEC with SRC
are negligible.

\begin{figure}[htb]
\begin{center}
{\includegraphics[width=10.cm]{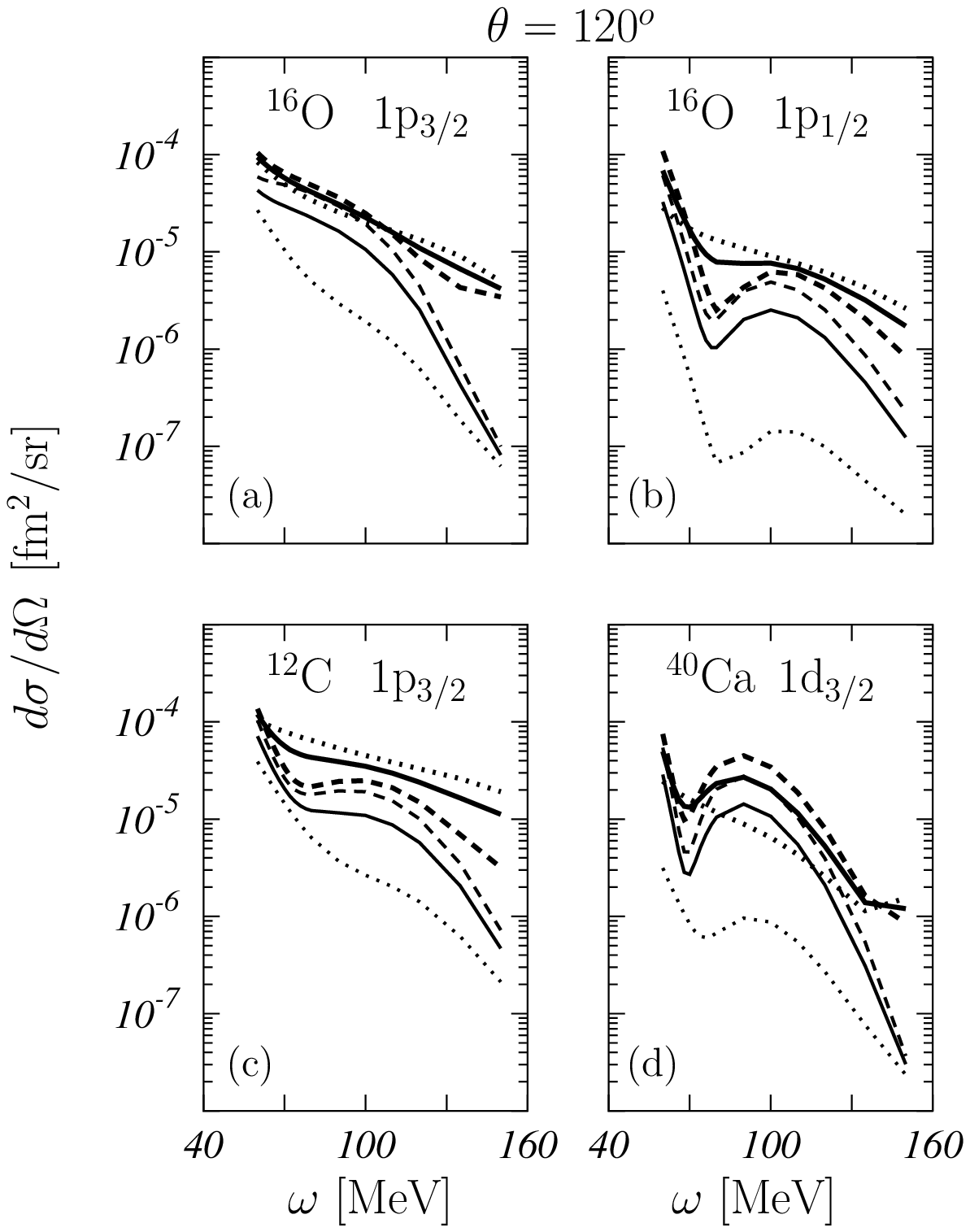}}
\end{center}
\vspace*{-0.6cm}
\caption{\small Cross sections as a function of the excitation energy
for a proton emission angle $\theta=120^o$.
The thin lines have been obtained by using OB currents only, and the
thick lines include MEC and SRC effects. 
The solid, dotted and dashed lines
represent the unpolarized cross section $\sigma$, $\sigma_{\parallel}$ and 
$\sigma_{\perp}$, respectively.}
\label{fig:CS-theta120}
\end{figure}

The behavior of the cross section at $\theta=120^o$ as a
function of the excitation energy, is shown in
Fig. \ref{fig:CS-theta120}. We have indicated with the full lines the
unpolarized cross sections $\sigma$, and with the dotted and dashed
lines the $\sigma_{\parallel}$ and $\sigma_{\perp}$, respectively. The
thin lines have been obtained by using OB currents only, while the
thicker lines include also MEC and SRC. It is interesting to notice
the large enhancement of the cross sections produced by the MEC in all
the cases we have considered.  These results are in contrast with
those of Bright and Cotanch \cite{bri93} which have calculated
unpolarized ($\gamma,p$) and ($\gamma,n$) cross sections for
$^{16}$O. We think that the difference is due to the description of
the FSI interaction. In the calculation of Bright and Cotanch a real
mean-field potential has been used.  We have verified that also in our
model, the use of real potential reduce the MEC effects.

\section{SUMMARY AND CONCLUSIONS}
\label{sec:conclusion}

We have investigated polarization observables, asymmetries, in
($\vec{\gamma},p$) reaction above the giant resonance region with a
model which considers the contribution of the SRC and includes MEC and
FSI effects. Our model has been applied to $^{12}$C, $^{16}$O and
$^{40}$Ca nuclei where we investigated the behavior of the
asymmetries for different values of the proton scattering angle and of
excitation energy.

In our investigation we found a relatively scarce sensitivity of the
asymmetries to the FSI. The sizes and the general behaviors of the
asymmetries are almost independent from the choice of the optical
potential. This was naively expected since the effects of the FSI
factorize to a large extent in the cross sections, therefore they
almost cancel in the asymmetries which are cross section ratios.
Our detailed analysis shows that the details of the angular
distributions depend on FSI, and they cannot be neglected. 

Therefore a proper description of this observable should be
done by using appropriate complex optical potentials. The importance
of FSI is well known in the case of the unpolarized cross sections,
but it is not so obvious for the asymmetries which are ratios of cross
sections. One would have expected in this last case, that the
dependence from the FSI would cancel out.

The SRC effects show up at large values of the emission angle. As in
the case of the unpolarized cross sections \cite{ang02}, theses effects
are obscured by the MEC. The presence of the MEC is certainly the
most important effect beyond the mean-field, one-body, description of
the process. We found that the largest MEC contributions are given by
the $\Delta$ currents. We found that these contributions are already
very important at energies well below the peak of the $\Delta$
resonance. We have shown that already at 80 MeV the shapes of the
asymmetries are strongly modified, at large emission angles, by the MEC.
We have also shown that, always at large emission angle, the cross
sections are strongly enhanced by the MEC.

While MEC effects have been clearly found in nuclear few-body systems,
they are not cleanly identified in medium-heavy nuclei. We have shown
that the asymmetries are extremely sensitive to the presence of MEC,
in particular to the $\Delta$ currents, which produce both
quantitatively and qualitative modifications of the angular
distributions. Measurements of this observable would provide clean
information about MEC in medium-heavy nuclear systems.

\section{ACKNOWLEDGMENTS}
This work has been partially supported by the agreement INFN-CICYT, by
the spanish DGI (FIS2005-03577) and by the MURST through the PRIN:
{\sl Teoria della struttura dei nuclei e della materia nucleare}.

%
%


\end{document}